\documentclass[conference]{IEEEtran}
\IEEEoverridecommandlockouts
\usepackage{cite}
\usepackage{amsmath,amssymb,amsfonts}
\usepackage{algorithmic}
\usepackage{graphicx}
\usepackage{textcomp}
\usepackage{tabularray}
\usepackage{xcolor}
\def\BibTeX{{\rm B\kern-.05em{\sc i\kern-.025em b}\kern-.08em
    T\kern-.1667em\lower.7ex\hbox{E}\kern-.125emX}}
\begin{document}

\title{Evaluating Predictive Models in Cybersecurity: A Comparative Analysis of Machine and Deep Learning Techniques for Threat Detection\\
}

\makeatletter
\newcommand{\linebreakand}{%
  \end{@IEEEauthorhalign}
  \hfill\mbox{}\par
  \mbox{}\hfill\begin{@IEEEauthorhalign}
}
\makeatother

\author{\IEEEauthorblockN{ Momen Hesham, Mohamed Essam, Mohamed Bahaa, Ahmed Mohamed, Mohamed Gomaa, Mena Hany, Wael Elsersy}
\IEEEauthorblockA{\textit{MSA University} \\
 Giza, Egypt \\
\{momen.hesham, mohamed.essam25, mohamed.bahaa4, ahmed.mohamed348, mohamed.gomaa2, mena.hany, wfarouk\}@msa.edu.eg }}

\IEEEoverridecommandlockouts
\IEEEpubid{\makebox[\columnwidth]{979-8-3503-6263-6/24/\$31.00 ©2024 IEEE \hfill}
\hspace{\columnsep}\makebox[\columnwidth]{ }}
\maketitle
\IEEEpubidadjcol

\begin{abstract}

As these attacks become more and more difficult to see, the need for the great hi-tech models that detect them is undeniable. This paper examines and compares various machine learning as well as deep learning models to choose the most suitable ones for detecting and fighting against cybersecurity risks. The two datasets are used in the study to assess models like Naive Bayes, SVM, Random Forest, and deep learning architectures, i.e., VGG16, in the context of accuracy, precision, recall, and F1-score. Analysis shows that Random Forest and Extra Trees do better in terms of accuracy though in different aspects of the dataset characteristics and types of threat. This research not only emphasizes the strengths and weaknesses of each predictive model but also addresses the difficulties associated with deploying such technologies in the real-world environment, such as data dependency and computational demands. The research findings are targeted at cybersecurity professionals to help them select appropriate predictive models and configure them to strengthen the security measures against cyber threats completely.

\end{abstract}
\begin{IEEEkeywords}
 Cybersecurity, Machine Learning, Deep Learning, Predictive Models, Ensemble Methods, Threat Detection, Model Evaluation, Artificial Neural Networks, SVM, Random Forest.
 \end{IEEEkeywords}

\section{Introduction}
The recent digital era does not only pose technical issues but also, it is a vital part of the national and international security, the global economy, and the data of personal information protection. With the deep merging of technology into the daily life of people through smart devices, cloud computing, and interdependent systems, there has been a dramatic growth in the possibilities of cyber-attacks. IDC expects that the number of connected devices to the internet will amount to 30 billion by the year 2025, and each of them may pose as a doorway for hackers \cite{b1}.

Due to the variety of cyber attacks (e.g. financial theft, corporate espionage, attacks against crucial power structures, and extensive data breaches), it is necessary to consider the matter very seriously. As per the Cybersecurity and Infrastructure Security Agency (CISA), the three most utilized kinds of cyber-attacks include phishing, ransomware, distributed denial-of-service (DDoS) attacks, and advanced persistent threats (APTs) thereby requiring specific solutions for cybersecurity detection and mitigation (CISA, 2019). The devastating impact of these crimes from a financial standpoint is colossal, with the cost of cybercrime hitting more than ten trillion dollars globally in 2025 alone, according to an estimate report \cite{b4}.

In the modern era, AI and ML make new inputs to cybersecurity, and now we have tools with such fabulous speed and efficiency to foresee and fight against cyber threats. A wide variety of technologies are capable of learning the behavior patterns over big data to reveal an outlier that might mean an occurrence of a security breach, according to the research. Nevertheless, these AI-based models fall under different forms of disparities which always depend on several conditions including the quality of data used, model architecture, and the exact thing they were made to discover.

This research will conduct a comparative analysis of different predictive models that have been used in cybersecurity among other implementations and test them exhaustively to identify the best-performing models. The sought-after output will be obtained through a thorough assessment of models on multiple grounds, accuracy, speed, adaptability to different types of cyber threats, and computational efficiency. Then, the significance of each model will be revealed. This approach is significant in understanding the models' real-world scenarios in terms of which finetuned settings provide the best blend of fast convergence, accuracy, and low cost.

Additionally, this research will touch on the complexities of implementing these models in the operations and the possible challenges related to privacy, the need for frequent retraining of the models with new data, and the security of AI systems from possible adversarial attacks. Henceforth, the presented study aims to remove these challenges and to provide a useful comparison of the efficiency of the various models, thus guiding cybersecurity specialists towards picking the best choice of tools and techniques that fit their specific needs.
\section{Related Work}
this study investigates the effectiveness of various machine learning models for intrusion detection in wireless sensor networks (WSNs) \cite{b5}. The methodology involved collecting a dataset of network traffic data, simulating cyber-attacks, and pre-processing the data for machine learning applications [1]. Several machine learning models, including Support Vector Machines (SVM), Random Forest (RF), and Artificial Neural Networks (ANN), were trained and evaluated on their intrusion detection accuracy \cite{b5}. The results indicated that the RF model achieved the highest accuracy in detecting cyber-attacks, followed by SVM and ANN \cite{b5}. However, the study acknowledges limitations such as the dependence on the specific dataset used and the potential for falsepositives\cite{b5}

\vspace{0.2cm}

This study \cite{b6} compares various machine learning techniques for cyber attack detection in smart grids. Their methodology involved collecting data from a simulated smart grid environment and injecting different cyber attacks [1]. Several machine learning models, including support vector machines (SVM), random forests (RF), and artificial neural networks (ANN), were trained and evaluated for their accuracy in detecting cyber attacks \cite{b6}. The results showed that ANN achieved the highest accuracy, followed by RF and SVM \cite{b6}. However, a limitation of this study is that it relied on simulated data, which may not perfectly reflect real-world scenarios \cite{b6}. Additionally, the effectiveness of the proposed methods might vary depending on the specific type of cyber attack \cite{b6}.

\vspace{0.2cm}

A recent study by Choras and Kozik (2023) investigated machine learning-based techniques for detecting cyber-attacks on web applications. Their methodology involved collecting a dataset of web traffic data, pre-processing the data to extract relevant features, and training various machine learning models to classify traffic as normal or malicious. The results indicated that the proposed approach could achieve high accuracy in cyber-attack detection. However, a limitation of the study is the reliance on a simulated dataset, which may not generalize well to real-world scenarios \cite{b7}.

\vspace{0.2cm}

The application of ER-VEC (Embedded Representation Vector Machine) for cyber-attack detection on the CICIDS2017 dataset offers a promising approach, but some limitations need to be addressed. While ER-VEC demonstrates effectiveness in anomaly detection \cite{b8}, its dependence on parameter tuning can influence accuracy. Additionally, the generalizability of ER-VEC to other intrusion detection datasets might require further investigation.
In their methodology, the authors likely pre-processed the CICIDS2017 data, which contains various network flow features, for compatibility with ER-VEC. This might have involved scaling or normalization techniques. They then employed ER-VEC to learn embedded representation vectors for the network flow data. The anomaly detection process would involve establishing a baseline for normal network traffic patterns based on the learned vectors. Deviations from this baseline could then be flagged as potential cyberattacks.

\vspace{0.2cm}

\cite{b17} in their article "Research on Cyber Attack Modeling and Attack Path Discovery" propose techniques] that will allow IT specialists to possibly identify the routes through which attacks occur and how they spread. This research work is oriented toward developing attack models which in turn can simulate attack circumstances and also derive potential attack vectors within a system. This knowledge becomes instrumental in the investigation of the logical sequence of the possible stages of an attacker that will, in turn, contribute to timely predictions and implementation of preventive strategies in cybersecurity methodology. The usefulness of the capability to build attacks and predict attack paths for more practical and efficient security measures cannot be overemphasized and serves to enhance the predictive function of cybersecurity systems. This method, essentially, meshes perfectly with the objectives of this experiment, which aspire to determine how different predictive models indeed detect and stave off cyber threats in different environments.

\vspace{0.2cm}

\cite{b18} carry out a study that is crucial in looking at the impact of cyber-physical threats on transmission systems by analyzing Zheng et al. (2021) work on " CYB-PHYTHREAT: Impact of Cyber-Physical Attack on Transmission Systems". This paper dwells we dual nature of threats—both cyber and physical—that modern power systems confront, being the latter highly sophisticated and advanced. The effort to protect critical infrastructure is therefore crucial. For the conduct of the study, this fictitious company will make use of the widely used Python scripting language for sensitive tasks and Pandapower for modeling the power grid. The analysis puts a major focus on critical attack scenarios such as various line topology changes, outages, and state estimation by using them to develop an effective defense against increasingly complex cyber-physical attacks. The results at weakness by Huang and his teammates stress the importance of more advanced modeling techniques in forecasting and managing such multiple threats. As a result, these kinds of findings might provide a foundation for the development of a framework for improving infrastructure resilience.

\vspace{0.2cm}

\cite{b19} address the vulnerabilities in healthcare IoT systems in their paper, "Availability Models for Healthcare IoT Systems: Category and Research And How Vulnerabilities Are Violated". The healthcare sector is considered as the area of study by the research and it reveals how IoT systems are vulnerable and discusses various types of cyber-attacks that IoT systems could be subjected to. By applying Markov models to assess the security and reliability of IoT infrastructures during their research, they were capable of providing valuable knowledge of the specific vulnerabilities and possible attacks of healthcare IoT systems. This function is particularly important in the light of growing healthcare digital technologies trend, indicating the critical requirement for strong security mechanisms to be in place to protect the dear health information from the emerging cyber dangers.

\vspace{0.2cm}
\cite{b20} "DDoS Attack Detection via Artificial Neural Networks," Jai Dalvi and his team examine the opportunity of using Artificial Neural Network (ANN) for detecting Distributed Denial of Service (DDoS) attacks, mainly involving network and transport layers, such as UDP-Flood, Smurf, HTTP-Flood, and SiDDoS. Dalvi's team worked to create a model to not only detect the occurrence of these attacks but also analyze the time and space complexity of the attack to improve the time efficiency of the detection process. The researchers highlight that ANNs can be well-performing models because of their accuracy in detecting and categorizing DDoS attacks and, as a result, support their prospect of being a reliable tool in cybersecurity defense systems. This study serves the purpose of our research as it is an example of the application of deep learning techniques in cyber security, which falls within the category of our examination of the predictive models of threat detection based on recent technological development.
\begin{table}
\centering
\caption{Results of previous works}
\label{tab: table1}
\begin{tblr}{
  width = \linewidth,
  colspec = {Q[179]Q[454]Q[292]},
  hlines,
  vlines,
}
Paper Ref & Models                                                                      & Results                                                                                                                       \\
         \cite{b5} & {RF\\GBM\\LGBM\\Catboost}                                                   & {98.5\% accuracy\\98.9\% accuracy\\99.3\% accuracy\\99\% accuracy}                                                            \\
         \cite{b7} & {MLP\\RFC\\SVC\\tree\\GNB\\SGDC\\KNC}                                       & {99.31\% accuracy\\96.95\% accuracy\\95.4\% accuracy\\93.04\% accuracy\\87.68\% accuracy\\85.00\% accuracy\\70.51\% accuracy} \\
         \cite{b17} & {Meta-Asset\\Meta-Attack\\ACO}                                              & Not mentioned                                                                                                                 \\
          \cite{b18}& {Line Switching Attack\\Generator and Line Outage\\State Estimation Attack} & Not mentioned                                                                                                                 \\
          \cite{b19}& Markov                                                                      & Not mentioned                                                                                                                 \\
         \cite{b20} & ANN                                                                         & 89.62\%                                                                                                                       
\end{tblr}
\end{table}
\vspace{5mm}
\section{methodology}

\subsection{Data Preprocessing}
The cybersecurity attacks dataset as well as the wireless sensor network dataset were carefully processed for the analysis of machine learning and deep learning models through meticulous data preprocessing. At first, columns that were excessive or repetitive were removed to optimize the datasets so that unnecessary computations were reduced and the input data maximized. Next, the one-hot encoding method was used to convert the categorical variables into a format that works with both traditional machine learning and neural networks. That way, the model will be compatible and usable for different learning approaches. The quality check was the next step including a check for missing values that may lead to bias or invalidation of the produced model. In addition to the feature scale disparity that may have an unequal effect on model performance, all numerical features were standardized to have zero mean and one variance. This step is commonly used in algorithms that assume data following the normal distribution. Data sets are consequently split into training and testing ones based on an 80-20 ratio to provide opportunities for model testing on new and unseen data, which in turn help us to assess and determine a model's generalizability and performance in real-life situations. Last step, imputation techniques were applied to fill in the blanks for any still missing values, using various imputation strategies tailored for different features and dataset structures. The careful preprocessing of the datasets not only allowed the predictive models, whether machine learning or deep neural networks, to function at optimized efficiency and accuracy, but also prepared the data for successful feature extraction and model training.
\subsection{Training models}
In this study, we employ several well-established machine learning models and deep learning models, each selected for its specific strengths in handling classification problems within the domain of cybersecurity

\subsubsection{Machine learning}
 The Naive Bayes classifier is used for its proficiency in handling large datasets and its effectiveness in making predictions based on the Bayes theorem, assuming independence between predictors. In cybersecurity, Naive Bayes is particularly useful for spam detection and malicious activity recognition due to its speed and efficiency in training and prediction, even with substantial data inputs \cite{b9}.
 \vspace{5mm}
\begin{equation*}
P(x_{i}\mid y) = \frac{1}{\sqrt{2\pi \sigma_y^{2}}} \exp \left(-\frac{(x_{i} -\mu_{y})^2}{2\sigma_y^{2}} \right)
\end{equation*}
\vspace{5mm}
Decision trees are utilized for their ability to create clear, understandable models that mimic human decision-making logic by branching out possible outcomes based on feature values. This model is adept at identifying attack patterns by categorizing data into branches based on feature characteristics, making it suitable for intrusion detection and preventing false positives in threat detection \cite{b10}.

Random Forest Building on the simplicity of decision trees, Random Forest incorporates ensemble learning to improve predictive accuracy and control over-fitting by averaging multiple decision trees trained on different parts of the dataset. This robustness makes Random Forest an excellent choice for detecting complex patterns in cybersecurity threats, where data often contains noise and non-linear relationships \cite{b11}.

 KNN is included for its straightforward, instance-based learning where the output is a class membership determined by a majority vote of its neighbors. It's highly effective in anomaly detection where the similarity between known cases and potential threats is critical for security assessments \cite{b12}.
 \vspace{5mm}
 \begin{equation*}
d(x,y)=\sqrt{\sum_{i=1}^{n}(Xi-Yi)^{2}}
\end{equation*}
\vspace{5mm}
 SVM is applied due to its capability to handle high-dimensional data and its effectiveness in classification by finding the hyperplane that best divides a dataset into classes. In cybersecurity, SVM is advantageous for identifying boundaries between normal operations and potential threats with high accuracy, especially useful in scenarios with unclear or overlapping threat classifications \cite{b13}.
 
\vspace{0.2cm}

Each model has its training and is specially selected to tackle specific cybersecurity challenges. For example, some of them can detect highly complex attack patterns with a high degree of accuracy, others can quickly classify threats. This enables the security system to be strengthened and its response strategies to be improved.
\subsubsection{Deep Learning}
Besides classical machine learning techniques, this work relies on the most advanced deep learning architectures which are known for their superior performance in image and pattern-based tasks. Also, these architectures become more and more relevant in the cybersecurity field where they are mainly used in automated event detection and response systems.

 VGG16 and VGG19 models which are well-known for being simple yet deep with 16 and 19 layers of convolutional layers, are used because they can discover intricate features of image data, like network traffic, which is similar to pattern acknowledgment. These models have proved very successful in finding small deviations in traffic data which can be an indicator of more advanced types of cyberattacks raising alarm and therefore, increasing the performance of security systems\cite{b14}.
 ResNet18, and ResNet50, which have the size of 18 and 50 layers, respectively, build on residual learning to mitigate problems in training very deep networks. This novel architecture is imperative for cybersecurity in that it is good at discerning complex attack patterns without the effect of vanishing gradients, often encountered when working with deep networks. In essence, ResNets are proficient in spotting such complex patterns across huge data volumes, which can help identify and counter the sophisticated and long-term forms of cyberattack\cite{b15}.
 The inception model, which is also called GoogleNet, has the concept that a network inside the network has several convolutional layers of different sizes simultaneously. This structure allows the model to track data at different levels simultaneously offering a detailed and layered understanding of the observed information. In the context of cybersecurity, Inception is noticeable for reconciling multi-tiered network data features, and it can perfectly be used in intrusion detection systems\cite{b16}.

\section{EXPERIMENT AND RESULTS }

\subsubsection{Dataset}
The research uses two specific data resources with a great data set that can help analysis of cybersecurity threats in multifaceted network conditions. The first dataset is known as the Cybersecurity Attacks Dataset ('cybersecurity-attacks.csv') which reflects the live network traffic patterns from the monitoring process in the IT infrastructure of multinationals over 6 months. The documents are made up of 40,000 records, and each of them comprises 25 attributes. These reports describe diverse traits of network traffic and cyber attacks. The distinctive features of traffic logging comprise timestamps, source and destination IP addresses, port numbers, used protocols, packet lengths, types, detailed attack signatures, and methods of handling network protection mechanisms. Furthermore, the dataset includes huge amounts of data from the log and alerts in the case of the firewall logs as well as IDS/IPS which finds its use in real-time analysis and capturing of the suspect activities. This dataset's formulation is indeed very appealing to those supervised learning algorithms that belong to the category of classifying and predicting cyber threats and rank first among the research-created ones.
The second Dataset is called the Wireless Sensor Network Dataset (WSN-DS.csv), which simulates an actual environment of a Wireless Sensor Network (WSN) with the data collected under different attack scenarios. This simulation is dedicated to diagnosing the network’s weaknesses and the behavior of the various sensors and nodes. The data set comprises 374661 records with 19 attributes that are interconnected for checking the operational status as well as the network health of the sensor nodes. Specific features that should be considered while developing software include IDs of the sensors, timestamps, whether the sensor is a cluster head or not, the distance to the cluster head, energy consumption, and data transmission metrics. Such parameters are of vital significance for evaluating the networks of sensor nodes vs. their vulnerability to cyber-attacks. These records are marked with a label that indicates normal operations or attack; this target label is vital for the development of predictive models designed to detect and respond to hypothetical threats that address similar ecosystems.
Through combining these datasets linked to our study we expect to create a clear picture of the office of network activities and attack paths across various technology ecosystems. The following section will add clarity to the previously mentioned assessment by evaluating the efficacy of different predictive models in terms of cybersecurity applications so that the developed models retain high integrity and the ability to detect subtle intimation of security threats.
\begin{figure}[h]
  \centering
  \vspace{5mm}
  \includegraphics[width=0.41\textwidth]{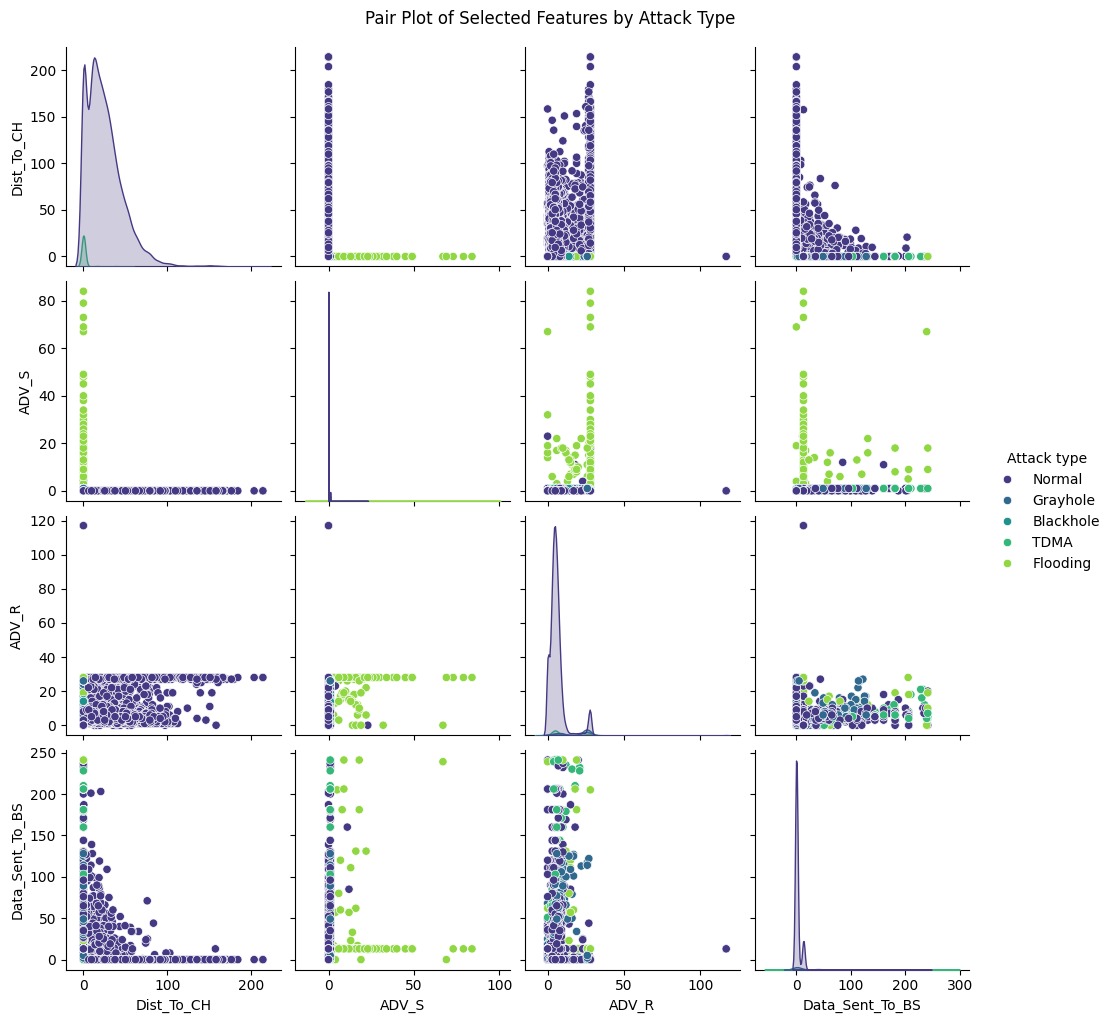}
  \caption{Pair Plot of Selected Features by Attack Type:}
  \label{fig2}
\end{figure}
\vspace{5mm}
\begin{figure}[h]
  \centering
  \vspace{5mm}
  \includegraphics[width=0.41\textwidth]{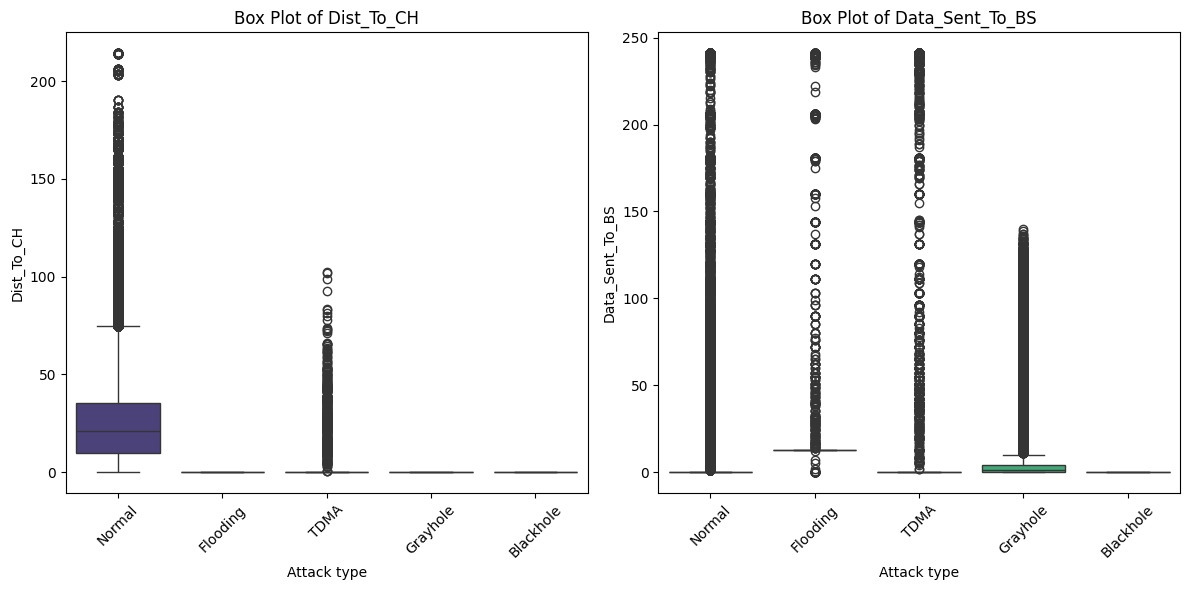}
  \caption{Box Plot Analysis of Distance to Cluster Head and Data Sent to Base Station by Attack Type}
  \label{fig2}
\end{figure}
\vspace{5mm}
\subsubsection{ Feature Selection}
Feature selection has a pivotal role in our study that enables the development of more accurate prediction models through the reduction of data clutter and avoiding overfitting. We start with identifying the highly informative features, for example, unusual traffic patterns on the network or specified protocol abuses.

As regards traditional machine learning models such as Decision Trees and SVM, we choose statistical methods including correlation matrices for removing duplicates and chi-squared tests for determining the most significant predictors. Thereby only the most important features are used and the model can be enhanced with higher efficiency and interpretability.

When using deep learning models like VGG16 and ResNet, the new goal is to prepare the data to fit the demands of these architectures. This stage includes image standardization, resizing, and augmentation to promote machine learning models to automatically recognize and extract features.

While modeling is in progress, the best features are chosen and constantly assessed with the help of performance metrics like accuracy and precision. This consecutive validation gives the possibility to amend the feature set to achieve the optimal models for the detection of cybersecurity threats.
\subsubsection{Model Execution}
The implementation of our concepts is the basis where the theory and practice begin to work together. As for the machine learning models covering Naive Bayes, Decision Trees, SVM, KNN, and Random Forest, the subsequent execution involves training the models on the preprocessed and feature-selected information. Models are configured in a way they can adjust their parameters targeting optimization via methods like grid search that determine which combinations of parameters are the most effective.

The VGG model, VGG16, VGG19, ResNet model, ResNet18, ResNet50, and Inception models are trained using the same dataset, but they take a different approach to overcome their intrinsic complexity. These models use the backpropagation technique for updating the weights and batch normalization which helps these models to have a mean output of zero and a variance of one during the training process. Learning rates are calibrated and strategies like the early stopping of training are used to prevent overfitting.

Both of the models are tested rigorously to identify their validity. To achieve this, we need to run the models on a separate testing set which was not exposed to the training set. By doing so, we can ensure the robustness of the results and their generalizability. Performance assessments that include accuracy, recall, precision, and F1-score are utilized to determine and compare the efficiency of each model in detecting and forecasting cyber security threats.

With the accurate building of these models, our goal is to come up with resilient systems that offer improved cyber-threat detection and prediction.

\begin{table}[ht]
    \centering
    \small
    \caption{first dataset Model performance metrics}
    \begin{tabular}{|l|c|c|c|c|}
        \hline
        Model Name & Accuracy & Precision & Recall & F1-score \\ \hline
        Naive Bayes & 84.4\% & 93.3\% & 84.8\% & 88.1\% \\ \hline
        Decision tree& 99.4\% & 99.4\% & 99.4\% & 99.4\% \\ \hline
        Random forest & 99.7\% & 99.7\% & 99.7\% & 99.6\% \\ \hline
        KNN & 98.8\% & 98.8\% & 98.1\% & 98.8\% \\ \hline
        SVM & 90.7\% & 82\% & 90.7\% & 96.2\% \\ \hline
        vgg16 & 99\% & 99\% & 89\% & 94\% \\ \hline
        vgg19 & 98\% & 95\% & 94\% & 95\% \\ \hline
        resnet18 & 97\% & 98\% & 38\% & 55\% \\ \hline
        resnet50& 97\% & 91\% & 38\% & 54\% \\ \hline
        inception& 98\% & 90\% & 99\% & 94\& \\ \hline
        Extra trees& 99.7\% & 99.7\% & 99.7\% & 99.7\% \\ \hline
    \end{tabular}
    \vspace{10mm}
    \label{tab:my-table}
\end{table}
\begin{table}[ht]
    \centering
    \small
    \caption{second dataset Model performance metrics}
    \begin{tabular}{|l|c|c|c|c|}
        \hline
        Model Name & Accuracy & Precision & Recall & F1-score \\ \hline
        Naive Bayes & 33\%& 34\% & 33\% & 33\% \\ \hline
        Decision tree& 35\%& 35\% & 36\% & 36\% \\ \hline
        Random forest & 36\% & 36\% & 39\% & 37\% \\ \hline
        KNN & 32\% & 32\% & 46\% & 38\% \\ \hline
        SVM & 34\% & 34\% & 39\% & 37\% \\ \hline
        vgg16 & 33\% & 33\% & 66\% & 44\% \\ \hline
        vgg19 & 34\% & 37\% & 10\% & 26\% \\ \hline
        resnet18 & 34\% & 34\% & 42\% & 37\% \\ \hline
        resnet50& 33\% & 33\% & 42\% & 37\% \\ \hline
        inception& 33\% & 33\%& 94\% & 49\% \\ \hline
        Extra trees& 34\% & 34\% & 38\% & 36\% \\ \hline
    \end{tabular}
    \vspace{10mm}
    \label{tab:my-table}
\end{table}
\section{DISCUSSION}
From this diverse analysis of different machine learning and deep learning models, we have come to know many interesting facts about their efficiency and the field of cybersecurity. In particular, the Random Forest and Extra Trees ensemble techniques have been shown to produce better results, suggesting their appropriateness when dealing with complex security challenges that contain noises and non-linear relations in the data which can be high dimensional. On the contrary, conventional models such as Naive Bayes and SVM have exhibited different performance patterns, which illustrate the necessity to be selective about the choice of model, taking into account the specific data features and threat conditions at hand. These outcomes demonstrate the ability of highly developed machine learning approaches to do not only the role of improving detection capabilities in threats but also the speed and accuracy in the response to cybersecurity. Nevertheless, the study outlines problems as well: namely, the reliance on large, well-prepared datasets for training and the high computational burden from the meticulous implementations of deep learning models. In addition, the issue of the applicability of these models to real-world situations is still a question mark because cyber threats in simulated environments may not completely reflect the dynamic nature of the real cyber environment. A primary research target going forward would thus be the enhancement of hybrid models, combining the interpretative features of traditional machine learning with the predictive power of deep learning, and their use in real-time cybersecurity systems mainly to increase adaptability to varying threats. The practical applicability and efficiency of AI-based cybersecurity tools can ultimately be long-term beneficiaries of this approach.

\section{CONCLUSION AND FUTURE WORK}
In this research, machine learning and deep learning models were evaluated through a holistic approach to evaluate these models through two given datasets which are very complex to deal with the cyber-security threats. This observation resulted in the fact that algorithmic models like Random Forest and Extra Trees demonstrated outstanding performance and showed a sufficient idea for protecting complex cybersecurity data. On the other hand, the Naive Bayes and SVM did not show the same level of performance. This underlines the need for the right model selection depending on specific data and particular types of threats. Meanwhile, despite these promising outcomes, the lack of large data sets put into order and the computational overhead of advanced models, which hamper their implementation globally, remain the main problems. Future research should concentrate on the creation of a hybrid model which is the simplicity and even interpretability of machine learning the advanced recognizing patterns capabilities of deep learning, to get the real-time applications in the various cyber security. Lastly, we need to develop adaptive learning models that adapt to evolving risks. They can provide innovative and solid solutions, which are needed in the rapid change of the cyber security landscape. It has therefore been made possible in this work to gain a good depth of understanding and implementation of AI in cybersecurity which calls for more effort and investment in bridging the theoretical research to practical, real-time, effective problems.

\vspace{12pt}
\color{red}

\end{document}